\documentstyle[preprint,prd,aps,epsf]{revtex}
\begin{document}
\tighten
\thispagestyle{empty}
\draft
\title{\hfill {\small DOE-ER-40757-102} \\
%\vspace*{-12pt}\hfill {\small UTEXAS-HEP-97-11} \\
%\vspace*{-12pt}\hfill {\small MSUHEP-70610} \\
\hfill {\small UTEXAS-HEP-97-11} \\
\hfill {\small MSUHEP-70610} \\
\hfill \\ 
Higgs-photon associated production at hadron colliders}

\author{Ali Abbasabadi$^1$, David Bowser-Chao$^2$, Duane A. Dicus$^3$ and 
Wayne W. Repko$^4$}
\address{$^1$Department of Physical Sciences, Ferris State University, 
Big Rapids, Michigan 49307}
\address{$^2$Department of Physics,
University of Illinois at Chicago, Chicago, Illinois 60607}
\address{$^3$Center for Particle Physics and Department of Physics 
University of Texas, Austin, Texas 78712}
\address{$^4$Department of Physics and Astronomy 
Michigan State University, East Lansing, Michigan 48824}

\date{\today}
\maketitle 
\begin{abstract}
We present cross sections for the reactions $p\bar{p}\rightarrow H\gamma$ and
$pp\rightarrow H\gamma$ arising from the subprocess $q\bar{q}\rightarrow
H\gamma$. The calculation includes the complete one-loop contribution from all
light quarks as well as the tree contributions from the light quarks and
$c$ and $b$ quarks.
These are the main sources of Higgs-photon associated production at hadron 
colliders. At Tevatron energies, the cross section is typically between 0.1fb 
and 1.0 fb for $80\,\mbox{\rm GeV}\leq m_H\leq 160\,\mbox{\rm GeV}$, while at 
LHC energies it exceeds about 2.0 fb over the same range of $m_H$. While
substantial, these cross sections are not large enough to produce a signal
which is discernable above the backgrounds from $q\bar{q}\to b\bar{b}\gamma$ 
and $gg\to b\bar{b}\gamma$. 
\end{abstract}

\section{Introduction}

    The production of intermediate mass Higgs bosons at hadron colliders arises
primarily from the gluon fusion process $gg\rightarrow H$, which is dominated by
the top quark loop \cite{hhg}. Depending on the collider energy, there can also 
be a substantial contribution from the gauge boson fusion processes
$W^+W^-,ZZ\rightarrow H$ \cite{fus}. For Higgs boson masses $m_H$ less than 
$2m_W$, detection of the Higgs via its dominant decay mode $H\rightarrow
b\bar{b}$ must contend with a large background from $gg\rightarrow b\bar{b}$.
The production of Higgs bosons in association with a gauge boson, specifically 
$Z,W^{\pm}$ \cite{bj,mk} or a photon \cite{bpr,ab-cdr}, is potentially helpful 
in dealing with these backgrounds, although at a significant cost in the rate.

    In this report, we examine the possibility of using Higgs-photon associated
production in hadron collisions as a means of studying properties of the Higgs 
boson. The main source of the $H\gamma$ final state is quark-antiquark
annihilation. Two-gluon annihilation, which accounts for single $H$ production,
is forbidden by Furry's theorem because the gluons are in a color singlet state.
For light quarks, the direct annihilation into $H\gamma$ is suppressed by the
ratio $m_q/m_W$ and is negligible. This means that one-loop electroweak
corrections involving $W$'s, $Z$'s and top quarks dominate the light quark
contribution to $q\bar{q}\rightarrow H \gamma$. The diagrams involved in the 
one-loop quark-antiquark annihilation calculation are similar to those 
encountered in the calculation of $e\bar{e}\rightarrow H\gamma$, 
which has been performed for the Standard Model \cite{bpr,ab-cdr} and
its supersymmetric generalization \cite{ddhr}. As in the case of
$\mu\bar{\mu}\to H\gamma$ \cite{lt,ab-cdr1}, there are also tree-level
contributions from $c\bar{c}$ and $b\bar{b}$ annihilation, which are enhanced
by their larger couplings but suppressed by parton distribution function 
effects.

In the next section, we outline the extension of our previous tree
\cite{ab-cdr1} and one-loop results \cite{ab-cdr} to $q\bar{q}\rightarrow
H\gamma$. This is followed by a discussion.

\section{Outline of the calculation}

\subsection{Tree-level amplitudes}

    The tree-level diagrams are illustrated in Fig.\,1 and the resulting
amplitude is
\begin{equation}
{\cal M}^{\rm tree}  = 
\frac{e_qgm_q}{2m_W}\left[-i\left(\frac{p_1\!\cdot\!\epsilon^*}{p_1\!\cdot\!k}
- \frac{p_2\!\cdot\!\epsilon^*}{p_2\!\cdot\!k}\right)\bar{v}(p_2)
u(p_1)\,+\,
\left(\frac{1}{2p_1\!\cdot\!k} + \frac{1}{2p_2\!\cdot\!k}\right)
\bar{v}(p_2)\sigma_{\mu\nu}k_{\nu}u(p_1)\,\epsilon^*_{\mu}\,\right]\,,
\end{equation}
where $p_1$ is the momentum of the quark, $p_2$ the momentum of the antiquark,
$k$ the momentum of the $\gamma$ and $\epsilon$ its polarization and $e_q$ is 
the quark charge in units of the proton charge. For massless quarks, only 
the helicity non-flip contributions from
the factors $\bar{v}(p_2)u(p_1)$ and $\bar{v}(p_2)\sigma_{\mu\nu}u(p_1)$ are
non-zero. Thus, we expect the helicity flip tree amplitudes to contain a factor
of order $m_q/E$ relative to the non-flip amplitudes. Explicitly, we
find,
\begin{equation} \label{treeamp}
{\cal M}^{\rm tree}_{\lambda_1\lambda_2\lambda_{\gamma}} = 
-i\frac{e_qgm_q}{\sqrt{\displaystyle 2}\,m_W}\left(\frac{1}{2p_1\!\cdot\!k} + 
\frac{1}{2p_2\!\cdot\!k}\right)\left\{
\begin{array}{lcl}
\sin\theta\left[\lambda_{\gamma}(2|{\bf p}|^2 - E\omega) + |{\bf p}|\omega
\right] &,& \lambda_1\lambda_2 = ++ \\
\sin\theta\left[\lambda_{\gamma}(2|{\bf p}|^2 - E\omega) - |{\bf p}|\omega
\right]&,& \lambda_1\lambda_2 = -- \\
m_q\,\omega(1 + \lambda_{\gamma}\cos\theta) &,& \lambda_1\lambda_2 = +- 
\\
m_q\,\omega(1 - \lambda_{\gamma}\cos\theta) &,& \lambda_1\lambda_2 = -+ 
\\
\end{array}
\right.\,,
\end{equation}
where $E$ is the quark energy in the center of mass, $|{\bf p}| = \sqrt{E^2 -
m_q^2}$, $\omega$ is the photon energy, $\theta$ is the photon scattering
angle and $\lambda_{\gamma} = \pm 1$ is the photon helicity. It can be seen 
that the helicity flip amplitudes have an additional factor of $m_q$.

\subsection{One-loop amplitudes}

    The one-loop amplitudes for $q\bar{q}\rightarrow H\gamma$ receive 
contributions from
pole diagrams involving virtual photon and $Z$ exchange and from various box
diagrams containing quarks, gauge bosons and/or Goldstone bosons. There are 
also double pole diagrams whose contribution vanishes. This is illustrated in 
Fig.\,1. The main difference between $e\bar{e}$ annihilation and $q\bar{q}$
annihilation occurs in the crossed box diagram in the last row of Fig.\,1. 
Since both members of a quark doublet are charged, there are additional crossed
box diagrams containing $W$'s. This contribution can be obtained from our $Z$ 
box results by merely changing the coupling and replacing $m_Z$ by $m_W$. 

    In the non-linear gauges we chose \cite{ab-cdr}, the full amplitude 
consists of four separately gauge invariant terms: a photon pole, a $Z$ pole,
$Z$ boxes and $W$ boxes. These amplitudes can be written as
\begin{eqnarray}\label{g}
{\cal M}^{\gamma}_{\rm pole} & = &\,\frac{\alpha^2m_W}{\sin\!\theta_W}\,
\bar{v}(p_2)\gamma_{\mu}u(p_1)\left(\frac{
\delta_{\mu\nu}k\!\cdot\!(p_1 + p_2) - k_{\mu}(p_1 + p_2)_{\nu}}{s}
\right)\hat{\epsilon}_{\nu}(k){\cal A}_{\gamma}(s)\,, \\ [4pt]
{\cal M}^{Z}_{\rm pole} & = &\,\frac{\alpha^2m_W}{\sin^3\!\theta_W}\,\bar{v}
(p_2)\gamma_{\mu}(v_q + \gamma_5)u(p_2)\left(\frac{\delta_{\mu\nu}k\!\cdot
\!(p_1 + p_2) - k_{\mu}(p_1 + p_2)_{\nu}}{(s - m_Z^2) + 
im_Z\Gamma_Z}\right)\hat{\epsilon}_{\nu}(k){\cal A}_{Z}(s)\,, 
\\ [4pt]
{\cal M}_{\rm box}^Z & = &-\frac{\alpha^2m_Z}{4\sin^3\!\theta_W\cos^3\!
\theta_W}\,\bar{v}(p_2)\gamma_{\mu}(v_q + \gamma_5)^2u(p_1)
\left\{\left[\delta_{\mu\nu}k\!\cdot\!p_1 - k_{\mu}(p_1)_{\nu}
\right]{\cal B}_Z(s,t,u)\right.
\nonumber \\
&   &+ \left.\left[\delta_{\mu\nu}k\!\cdot\!p_2 - k_{\mu}(p_2)_{\nu}\right]
{\cal B}_Z(s,u,t)\right\}\hat{\epsilon}_{\nu}(k)\;, \\ [4pt]
\label{w}{\cal M}_{\rm box}^{W} & = &\,\frac{\alpha^2m_W}{2\sin^3\!\theta_W}\,
\bar{v}(p_2)\gamma_{\mu}(1 + \gamma_5)^2u(p_1) 
\left\{\left[\delta_{\mu\nu}k\!\cdot\!p_1 - k_{\mu}(p_1)_{\nu}
\right]{\cal B}_W(s,t,u)\right.\nonumber \\
&   &+ \left.\left[\delta_{\mu\nu}k\!\cdot\!p_2 - k_{\mu}(p_2)_{\nu}\right]
{\cal B}_W(s,u,t)\right\}
\hat{\epsilon}_{\nu}(k)\;,
\end{eqnarray}
where $s = -(p_1 + p_2)^2,t = -(k + p_1)^2\,$ and $u = -(k + p_2)^2$. Here,
$v_q$ denotes the $q\bar{q}Z$ vector coupling constant, $v_q = 1 -
4|e_q|\sin^2\!\theta_W$. In terms of the scalar functions defined in the 
appendices of our previous paper \cite{ab-cdr}, we have \cite{two}
\begin{eqnarray} \label{gam}
{\cal A}_{\gamma}(s) & = & -e_q\left\{4(6 + \frac{m_H^2}{m_W^2})
C_{23}(s,m_H^2,m_W^2) - 16C_0(s,m_H^2,m_W^2)\right. \nonumber \\ 
&   &\left.-\,\frac{16}{3}\frac{m_t^2}{m_W^2}\left(4C_{23}(s,m_H^2,m_t^2) 
 -  C_0(s,m_H^2,m_t^2)\right)\right\}\,,\\ [6pt]
\label{z}{\cal A}_Z(s) & = &\; -2I_3\left\{\left(5 - \tan^2\theta_W
 + \frac{m_H^2}{2m_W^2}(1 - \tan^2\theta_W)\right)C_{23}(s,m_H^2,m_W^2)\right.
 \nonumber \\
&   &\left.+ \left(\tan^2\theta_W - 3\right)C_0(s,m_H^2,m_W^2)\right. \nonumber
\\
&   &\left. - \frac{1}{2}\frac{m^2_t}{m^2_W}
\frac{1 - (8/3)\sin^2\theta_W}{\cos^2\!\theta_W}\left(4C_{23}(s,m_H^2,m_t^2) - 
C_0(s,m_H^2,m_t^2)\right)\right\}\,, \\
{\cal B}_Z(s,t,u) & = & -\tilde{e}_qA(s,t,u)\,, \\
{\cal B}_W(s,t,u) & = & -2I_3\left[A_1(s,t,u) + A_2(s,u,t)\right]
+ \tilde{e}_qA^{\prime}(s,t,u)\,,
\end{eqnarray}
with $I_3$ denoting the third component of the {\em external} quark weak 
isospin, $m_t$ being the top quark mass and the prime denoting the replacement
$m_Z\rightarrow m_W$. $\tilde{e}_q$ is the charge of the {\em internal} quark in
units of the proton charge. In this case, it is the helicity flip
contributions from the factors $\bar{v}(p_2)\gamma_{\mu}u(p_1)$ and 
$\bar{v}(p_2)\gamma_{\mu}\gamma_5 u(p_1)$ which survive in the
$m_q\rightarrow 0$ limit. This can be seen by noting that, in the center of
mass, we have
\begin{mathletters}
\label{loopamp}
\begin{eqnarray}
\bar{v}_+(p_2)\gamma_{\nu}u_+(p_1) & = & im_q(p_2 - p_1)_{\nu}/|{\bf p}|
\,,\\
\bar{v}_+(p_2)\gamma_{\nu}\gamma_5 u_+(p_1) & = & im_q(p_2 + p_1)_{\nu}
/E \,,\\
\bar{v}_+(p_2)\gamma_{\nu}u_-(p_1) & = & -2\sqrt{2}iE\xi^{(-)}_{\nu} \,,\\
\bar{v}_+(p_2)\gamma_{\nu}\gamma_5u_-(p_1) & = & -2\sqrt{2}i|{\bf p}|
\xi^{(-)}_{\nu} \,,
\end{eqnarray}
\end{mathletters}
with $\xi^{(-)}_{\nu} = (1,-i,0,0)/\sqrt{2}$.
If we define ${\cal M}^{\rm loop}$ as
\begin{equation} \label{mloop}
{\cal M}^{\rm loop} = {\cal M}^{\gamma}_{\rm pole} + {\cal M}^{Z}_{\rm pole} + 
{\cal M}^Z_{\rm box} + {\cal M}^W_{\rm box}\,,
\end{equation}
then ${\cal M}^{\rm loop}$ is predominantly helicity flip.

\subsection{Differential cross section}

    The differential cross section $d\sigma(q\bar{q}\rightarrow H\gamma)
/d\Omega_{\gamma}$ is
\begin{equation} \label{dsig}
\frac{d\sigma(q\bar{q}\rightarrow H\gamma)}{d\Omega_{\gamma}} =
\frac{1}{256\pi^2}\frac{s - m_H^2}{s^2}\sum_{\rm helicity}|{\cal M}|^2\,,
\end{equation}
where the invariant amplitude ${\cal M} = {\cal M}^{\rm tree} + {\cal M}^{\rm
loop}$. Because of the helicity structure discussed above, the interference
terms turn out to be negligible and one can simply add the tree and one-loop
cross sections.

To obtain the collider cross sections, we convolute Eq.\,(\ref{dsig}) 
with the appropriate quark and antiquark distribution functions using
\begin{equation}
\frac{d\sigma}{dm^2_{H\gamma}} = \frac{1}{s}\int_{\tau}^1\frac{dx}{x}f_q(x)
f_{\bar{q}}(\tau/x)\int_{-z_0}^{z_1}dz\frac{d\sigma(q\bar{q}\rightarrow H
\gamma)}{dz}\,.
\end{equation}
Here, $\tau = m^2_{H\gamma}/s$ , $z = \cos\theta_{\gamma}$ and $z_0$ and $z_1$ 
are determined by the choice of the rapidity and transverse momentum cuts. We 
used CTEQ3-1M parton distribution functions \cite{cteq} and imposed a rapidity
cut of 2.5 and a transverse momentum cut of 10 GeV on both the Higgs boson and 
the photon.

\section{Discussion}

    The total cross section for $p\bar{p}(pp)\rightarrow H\gamma$ is illustrated
in Fig.\,2 for an upgraded Tevatron energy of 2.0 TeV (left panel) and an 
LHC energy of 14 TeV (right panel). In each case, we show the tree 
contributions from light quarks ($u,d,s$), light quarks plus the $c$ quark
and light quarks plus $c$ and $b$ quarks. In addition, the loop contribution
from the light quarks is shown. The loop contributions from the $c$ and $b$
quarks are negligible. As asserted in the Introduction, Fig.\,2 confirms that 
the light quark tree contribution can always be neglected. 
Interestingly, the largest tree contribution is
due to the $c$ quark for both the Tevatron and the LHC, and the $b$ quark
contribution is always important. At the Tevatron energy, the loop diagrams
contribute about as much as the tree diagrams, while, for the LHC, the tree
diagrams dominate.

In our calculation of the heavy quark tree contributions, we simply convoluted
the cross section for $Q\bar{Q}\to H\gamma$ with the complete heavy parton
distribution functions \cite{cteq}, instead of subtracting from them the lowest
order logarithmic correction to avoid potential double counting of higher 
order gluonic contributions \cite{bhs,oltung}. 
Based on an earlier study of Higgs boson production
at the SSC \cite{dicwill}, this procedure is likely to somewhat overestimate
the heavy quark fusion contribution. The effects of QCD corrections have not 
been considered, although some of these, in particular the $K$ factor, usually
increase the heavy quark cross sections \cite{dgv}. We have not included top 
quark fusion in our results since the top quark is almost certainly too 
massive to be treated as a parton at LHC energies.

Although there are non-negligible numbers of events for a Tevatron 
with an upgraded luminosity of 20 - 30fb$^{-1}$/year \cite{tev33} 
or the LHC with a design luminosity of 100fb$^{-1}$/year, the
background to the signal $p\bar{p}(pp)\to H\gamma\to b\bar{b}\gamma$
must be considered. To do this, we have calculated the direct production of the
$b\bar{b}\gamma$ final state arising from the channels $p\bar{p}(pp)\to 
q\bar{q}\to b\bar{b}\gamma$ \cite{eemmg} and $p\bar{p}(pp)\to gg\to 
b\bar{b}\gamma$. In Table I, the 
background contributions are shown for cuts on the $b\bar{b}$ invariant mass 
$m_{b\bar{b}}$ in the vicinity of $m_H = 100\,$GeV and 200 GeV
and rapidity cuts $y_{\rm max} = 1.0$ and 2.5 on the $b$, $\bar{b}$ and 
$\gamma$. In addition to these cuts, we require the separation $\Delta R$ 
between the $\gamma$ and the $b$ and the $\gamma$ and the $\bar{b}$ to be 
greater than 0.4, the transverse momenta of the $b$, $\bar{b}$ and $\gamma$ to 
be greater than 15 GeV, and the $b\bar{b}\gamma$ invariant mass to be greater 
than 170 GeV. The latter two cuts reduce the background in the $m_H$ = 100 GeV 
region by about a factor of 5. They reduce the loop signal by about 40\% at
$\sqrt{s}$ = 1.8 TeV and 14 TeV, but the corresponding reduction in the tree
contribution is a factor of 15 at 1.8 TeV and a factor of 6.8 at 14 TeV. Thus
the tree contribution, though substantial, is difficult to isolate.

The background is compared to the signal in Table II for Higgs boson masses of 
100 GeV and 200 GeV. Despite the application of numerous cuts, it appears 
that the background is prohibitively large for the observation of a Higgs 
boson with Standard Model couplings produced in association with a photon.

Finally, to determine the sensitivity of associated production to changes in the
$t\bar{t}H$ coupling, we computed the cross section including a factor $\lambda$
multiplying the Standard Model coupling \cite{hks}. For a 2 TeV Tevatron 
upgrade, this effect is illustrated in Fig.\,3 as a function of $\lambda$.
The most obvious effect is the uniform decrease in the
cross section as $\lambda$ varies from $\lambda = 0$ to the Standard Model value
at $\lambda = 1.0$. With increasing $\lambda$, the cross sections eventually
exceed the Standard Model result. The enhancement in the cross section for
larger values of $\lambda$ is still not sufficient to
produce favorable $S/\sqrt{B}$ value. For example, for $\lambda = 10$ and $m_H =
100\,$ GeV, the cuts used for Table II give $S/\sqrt{B} \sim .22$. It is clear 
that any anomalous $H\gamma$ coupling must result in a cross section of several 
femtobarns and a fairly isotropic angular distribution in order to produce a 
signal that will survive the cuts required to reduce the $b\bar{b}\gamma$ 
background.

\acknowledgements

We would like to thank G. Kane for several helpful conversations and Wu-Ki Tung
for informative discussions about heavy quark parton distribution functions.
This research was supported in part by the U.S. Department of Energy under
Contract No. DE-FG013-93ER40757 and in part by the National Science Foundation 
under Grant No. PHY-93-07980.

\newpage

\begin{table}
\begin{center}
\begin{tabular}{ccccc}
\mbox{\rule[-6pt]{0pt}{18pt}}$\sqrt{s}$ & $y_{\rm max}$ &$98.5 < m_{b\bar{b}} <
101.5$& $99 < m_{b\bar{b}} < 101$ &
$198.5 < m_{b\bar{b}} < 201.5 $\\
\hline
1.8 TeV &2.5& 41.9 fb & 28.7 fb  & 20.0 fb \\
1.8 TeV &1.0& 5.57 fb & 3.73 fb  & 4.35 fb \\
14 TeV  &2.5& 340  fb & 227  fb  & 276  fb \\
14 TeV  &1.0& 29.7 fb & 19.8 fb  & 25.4 fb \\
\end{tabular}
\end{center}
\caption{Cross sections for the backgrounds $p\bar{p}\,(pp)\protect\to
q\bar{q}\protect\to\gamma b\bar{b}$ and $p\bar{p}\,(pp)\protect\to gg
\protect\to\gamma b\bar{b}$ are given for several cuts on the $b\bar{b}$ 
invariant mass $m_{b\bar{b}}$ (in GeV), a cut $y_{\rm max}$ on the rapidities 
of the $\gamma$, the $b$ and the $\bar{b}$, and other cuts discussed in the 
text.}

\end{table}

\begin{table}
\begin{center}
\begin{tabular}{cccccc}
\mbox{\rule[-6pt]{0pt}{18pt}}$\sqrt{s}$ &$y_{\rm max}$ & $\sigma(m_H = 100\;
{\rm GeV})$ & $S/\sqrt{B}$ &$\sigma(m_H = 200\;{\rm GeV})$& $S/\sqrt{B}$ \\
\hline
1.8 TeV &2.5&$4.41\times 10^{-2}$\,fb & .05 & $9.01\times 10^{-3}$\,fb&  .01 \\
1.8 TeV &1.0&$1.79\times 10^{-2}$\,fb & .05 & $4.57\times 10^{-3}$\,fb&  .02 \\
14 TeV  &2.5& 1.22\,\,fb              & .66 & $9.13\times 10^{-1}$\,fb&  .55 \\
14 TeV  &1.0&$2.43\times 10^{-1}$\,fb & .46 & $2.96\times 10^{-1}$\,fb&  .59 \\
\end{tabular}
\end{center}
\caption{The cross sections for the associated production of 100 GeV and 
200 GeV Higgs bosons are shown together with the ratio of the signal to the 
square root of the background ($S/\protect\sqrt{B}$) for the Tevatron and the
LHC. A luminosity of 50 fb$^{-1}$ is assumed at $\protect\sqrt{s} = 1.8\,$ TeV
and 100 fb$^{-1}$ at $\protect\sqrt{s} = 14\,$ TeV. The cuts used are discussed
in the text.}
\end{table}

\begin{figure}[h]
\hspace*{1.0in}
\epsfysize=2.0in \epsfbox{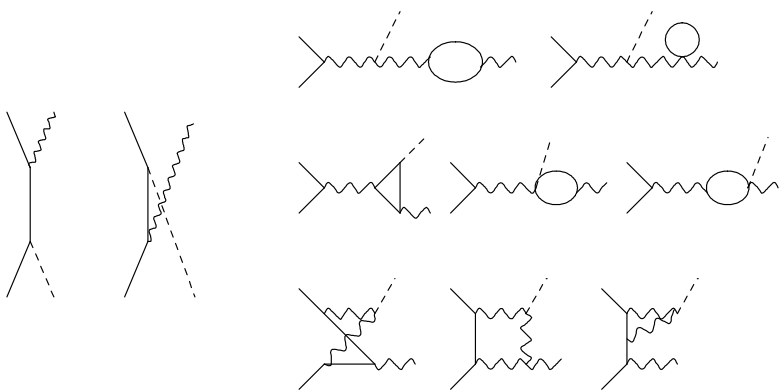}
\vspace{.10in}
\caption{Typical diagrams for the tree and one-loop contributions are shown.
 An external solid line represents a quark, a wavy
line a gauge boson, a dashed line a Higgs boson and an internal solid line a
quark, gauge boson, Goldsone boson or ghost.}
\end{figure}

\newpage

\begin{figure}[h]
\hspace*{.1in}
\epsfysize=2.75in \epsfbox{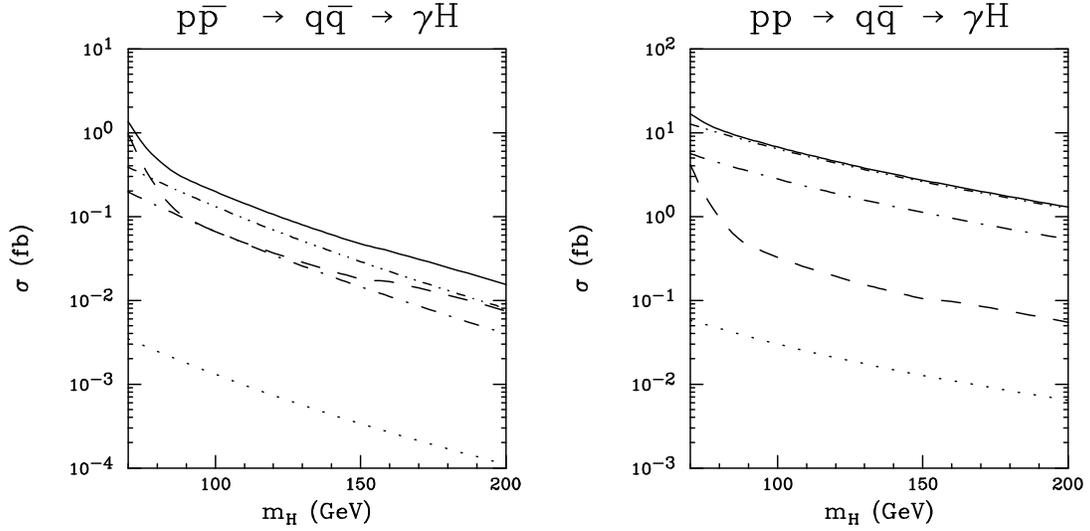}
\vspace{.10in}
\caption{The total cross section for $H\gamma$ production from $q\bar{q}$
annihilation in $p\bar{p}$ and $pp$ scattering is shown for a Tevatron energy of
2 TeV and and LHC energy of 14 TeV. In each panel, the dotted line is the light
quark tree contribution, the dot-dashed line is the light quark plus $c$ quark
tree contribution, the dot-dot-dashed line is the light quark plus $c$ and $b$
tree contribution and the dashed line is the loop contribution. The solid line 
is the sum of the tree and loop contributions. The cuts used are discussed in
the text.}
\end{figure}

\newpage

\begin{figure}[h]
\hspace*{1.25in}
\epsfysize=2.75in \epsfbox{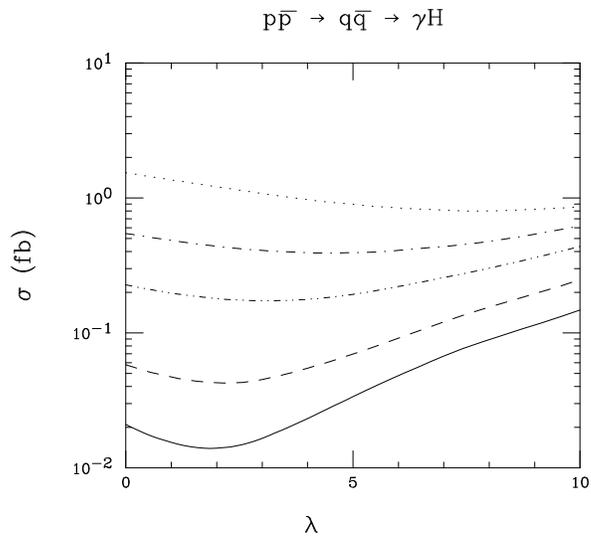}
\vspace{.10in}
\caption{The total cross section (tree + loop) for $H\gamma$ production from 
$q\bar{q}$ annihilation in $p\bar{p}$ scattering is shown as a function of the
$t\bar{t}H$ coupling (in multiples $\lambda$ of the Standard Model coupling)
at an upgraded Tevatron for several values of $m_H$. The dotted line is
$m_H = 70$\,GeV, the dot-dashed line $m_H = 80$\,GeV, the dot-dot-dashed line 
$m_H = 100$\, GeV, the dashed line $m_H = 150$\,GeV, and the solid line 
$m_H = 200$\,GeV.}
\end{figure}

\end{document}